\documentclass[english,aps]{revtex4}
\usepackage[T1]{fontenc}
\usepackage[latin9]{inputenc}
\usepackage{float}
\usepackage{graphicx}

\makeatletter

\providecommand{\LyX}{L\kern-.1667em\lower.25em\hbox{Y}\kern-.125emX\@}
\floatstyle{ruled}
\newfloat{algorithm}{tbp}{loa}
\floatname{algorithm}{Algorithm}

\@ifundefined{textcolor}{}
{%
 \definecolor{BLACK}{gray}{0}
 \definecolor{WHITE}{gray}{1}
 \definecolor{RED}{rgb}{1,0,0}
 \definecolor{GREEN}{rgb}{0,1,0}
 \definecolor{BLUE}{rgb}{0,0,1}
 \definecolor{CYAN}{cmyk}{1,0,0,0}
 \definecolor{MAGENTA}{cmyk}{0,1,0,0}
 \definecolor{YELLOW}{cmyk}{0,0,1,0}
 }

\renewcommand\[{\begin{equation}}
\renewcommand\]{\end{equation}}

\makeatother

\usepackage{babel}

\begin{document}

\preprint{This line only printed with preprint option}

\title{Computing the energy of a water molecule using MultiDeterminants:
A simple, efficient algorithm}

\author{Bryan K. Clark}

\email{bclark@princeton.edu}

\affiliation{Princeton Center For Theoretical Science, Princeton University, Princeton,
NJ 08544\\
Department of Physics, Joseph Henry Laboratories, Princeton University,
Princeton, NJ 08544}

\author{Miguel A. Morales}

\email{moralessilva2@llnl.gov}

\affiliation{Lawrence Livermore National Laboratory, 7000 East Avenue, Livermore, California 94550, U.S.A.}



\author{Jeremy McMinis}
\email{jmcminis@illinois.edu}

\affiliation{Department of Physics, University of Illinois at Urbana Champaign,
 Urbana, IL 61801}

\author{Jeongnim Kim}
\email{jnkim@illinois.edu}
\affiliation{National Center for Supercomputing Applications, University of Illinois at Urbana-Champaign, Urbana, IL 61801}

\author{Gustavo E. Scuseria}
\email{guscus@rice.edu}
\affiliation{Department of Chemistry and Department of Physics \& Astronomy, Rice University, Houston, TX 77005-1892, USA}

\begin{abstract}
Quantum Monte Carlo (QMC) methods such as variational Monte Carlo
and fixed node diffusion Monte Carlo depend heavily on the quality
of the trial wave function. Although Slater-Jastrow wave functions
are the most commonly used variational ansatz in electronic structure,
more sophisticated wave-functions are critical to ascertaining new
physics. One such wave function is the multiSlater-Jastrow wave function
which consists of a Jastrow function multiplied by the sum of Slater
determinants. In this paper we describe a method for working with
these wavefunctions in QMC codes that is easy to implement, efficient
both in computational speed as well as memory, and easily parallelized.
The computational cost scales quadratically with particle number 
making this scaling no worse than the single determinant case
and linear with the total number of excitations.
Additionally we implement this method and use it to compute 
the ground state energy of a water molecule.
\end{abstract}
\maketitle

\section{introduction}

Being able to accurately calculate the material properties of molecules
and solids is important for a variety of fields. There exist a variety
of different methodologies to accomplish this including density functional
theory \cite{martin2004electronic}, quantum chemistry \cite{szabo1996modern} 
and quantum Monte Carlo \cite{foulkes2001quantum}. All these methods
have different regimes of applicability and different tradeoffs between
accuracy and speed. For larger and more complicated systems, quantum
Monte Carlo methods are often a good choice due to their polynomial
scaling with particle number and high levels of accuracy. For ground
state calculations, the most commonly used QMC method is fixed node
diffusion Monte Carlo (FNDMC). FNDMC takes as input a trial wave function
$\Psi_{T}$ and returns the ground state energy of the fixed node
wave-function $\Psi_{FN}$, the wave-function with the lowest ground
state energy with the same nodes as $\Psi_{T}$ (i.e. $\Psi_{T}(R)=0\iff\Psi_{FN}(R)=0$).
The accuracy of FNDMC is dependent on the quality of the trial wave
function. 

Although the most commonly used form of a trial-wavefunction is the
Slater-Jastrow form, acheiving higher levels of accuracy require the
use of more sophisticated ansatz. This is especially true in intrinsically
multi-reference problems. Such ansatz are only useful if they can
be used with limited computational complexity and memory. A natural
extension of the Slater-Jastrow wave-function is the multiSlater-Jastrow
form. One  advantage of this wavefunction is that any state can be
represented  in the limit of a large 
enough number of determinants. Therefore, an answer
can be systematically improved upon by including more determinants
in the ansatz. In this work, we describe a new algorithm that allows
FNDMC to use the multiSlater-Jastrow form in an efficient manner.
To use a wave-function in FNDMC, there are two important and computationally
demanding requirements: the ability to compute ratios of the wavefunction
evaluated in two configurations that differ by the location of a single
particle and the ability to evaluate gradients and laplacians of the
wavefunction. Here we will describe algorithms to accomplish
both these tasks. These algorithms will require a minimal amount of
additional memory with respect to the single determinant case and
will scale (for a single step) with a computational complexity that goes as $O(n^{2}+n_{s}n+n_{e})$
where $n$ is the total number of particles, $n_{s}$  the total number of single excitations and $n_{e}$
is the total number of excitations. This scaling is the same with respect
to particle number as the single determinant case and scales linearly with 
the total number of excitations.  Other recent work has also proposed
an algorithm for computing ratios of multi-determinants \cite{nukala2009fast}.
The method described here scales better by a factor of the particle
number $n$ in almost all regimes of interest. In addition, it is
significantly simpler requiring less book-keeping and eliminating
the need for recursive trees. This simple structure also makes it
transparent that the computation involved can be efficiently parallelized.
In section II we more concretely introduce the wave function and the
necessary operations that must be performed. In Section III we describe
an efficient way of performing one of these operations: evaluating
wave function ratios. In section IV we focus on computing gradients
and laplacians of these terms. Section V discusses other important
implementation considerations. In section VI we use our method,
implemented in the code QMCPACK \cite{QMCPACK},
on the water molecule. Finally section VII summarizes our results.

\section{Background}

Fixed Node Diffusion Monte Carlo (FNDMC) take as input a variational
wave function. In this paper we discuss the multiSlater-Jastrow
which takes the form \[
\Psi=\exp[-J(R)]\sum_{k}\alpha_{k}\det M_{\uparrow k}\det M_{\downarrow k}\]
 where $J$ is a Jastrow function and $\alpha_{k}$ the weight of
the $k'$th determinant. The $(i,j)$ element of matrix $M_{k}$ is
equal to \[
M_{\sigma k(i,j)}=\phi_{j}(r_{i})\]
 where $\phi_{i}$ are 3 dimensional single particle orbitals (s.p.o)
selected from the set $O\equiv\{\phi_{1}...\phi_{\{n+m\}}\}$ where
$n$ is the total number of electrons and $m$ is the number of {}``virtual
orbitals.'' Although each matrix $M_{\sigma k}$ may, in principle,
contain an arbitrary set of orbitals selected from $O$, we consider
the typical case where the matrices differ by the action of particle-hole
excitations on a reference determinant (i.e. matrices that show up
in the multideterminant expansion differ from the reference determinant
by at most $K$ columns.) Replacing one (two, three) orbitals in the
reference determinant is respectively called a single (double, triple)
excitation. An example where such an expansion naturally arises is
the configuration interaction approach in quantum chemistry. 

By convention, notate the orbitals in the reference determinant as
$\phi_{1}...\phi_{n}$. The particle-hole excitations are then created
by replacing orbitals from the reference determinant with orbitals
selected from the $m$ virtual orbitals. Notate any orbitals in the
reference determinant that might be replaced to fill a matrix $M_{\sigma k}$
for $k>0$ as the ground state orbitals. The other orbitals in the
reference determinant (typically called frozen or core orbitals) 
will therefore show up in all determinants in
the expansion and can be ignored. We should note that the
canonical Slater-Jastrow form simply takes $\alpha_{k}=0$ for all
$k>0$. 

In quantum Monte Carlo calculations, particles are moved one at a time. 
We will call the movement of each particle a step. The movement of every 
particle once is then considered a sweep.  
There are two basic operations involving the wave-function in FNDMC.
They each have to be evaluated once per step (although for the laplacian
term, one can also evaluate it once per sweep for every particle $i$).  Unless 
explicitly stated otherwise, the computational complexities discussed in this paper
will be the cost per step. 
The two operations are evaluating the ratio of two wavefunctions that differ by
the location of a a single cooridinate \[
\frac{\Psi_{T}(r_{1}....r_{i}'...r_{n})}{\Psi_{T}(r_{1}...r_{i}....r_{n})}\]
 and evaluating the quantities \[
\frac{\nabla_{i}\Psi_{T}}{\Psi_{T}},\frac{\nabla^{2}\Psi_{T}}{\Psi_{T}}\]

We recall that the wave-function decomposes into the product of two
pieces: the Jastrow factor and the multideterminant sum. For computing
these quantities, it is sufficient to be able to individually compute
the properties for the Jastrows and the Multideterminant expansions.
Throughout the rest of this paper, we will focus on evaluating these
quantities for the sum of Slater determinants. As the Jastrow factor
terms can be dealt with using standard methods, we will not discuss
them further.

\section{Evaluating Wave Function Ratios}

\begin{figure}
\includegraphics[scale=0.5]{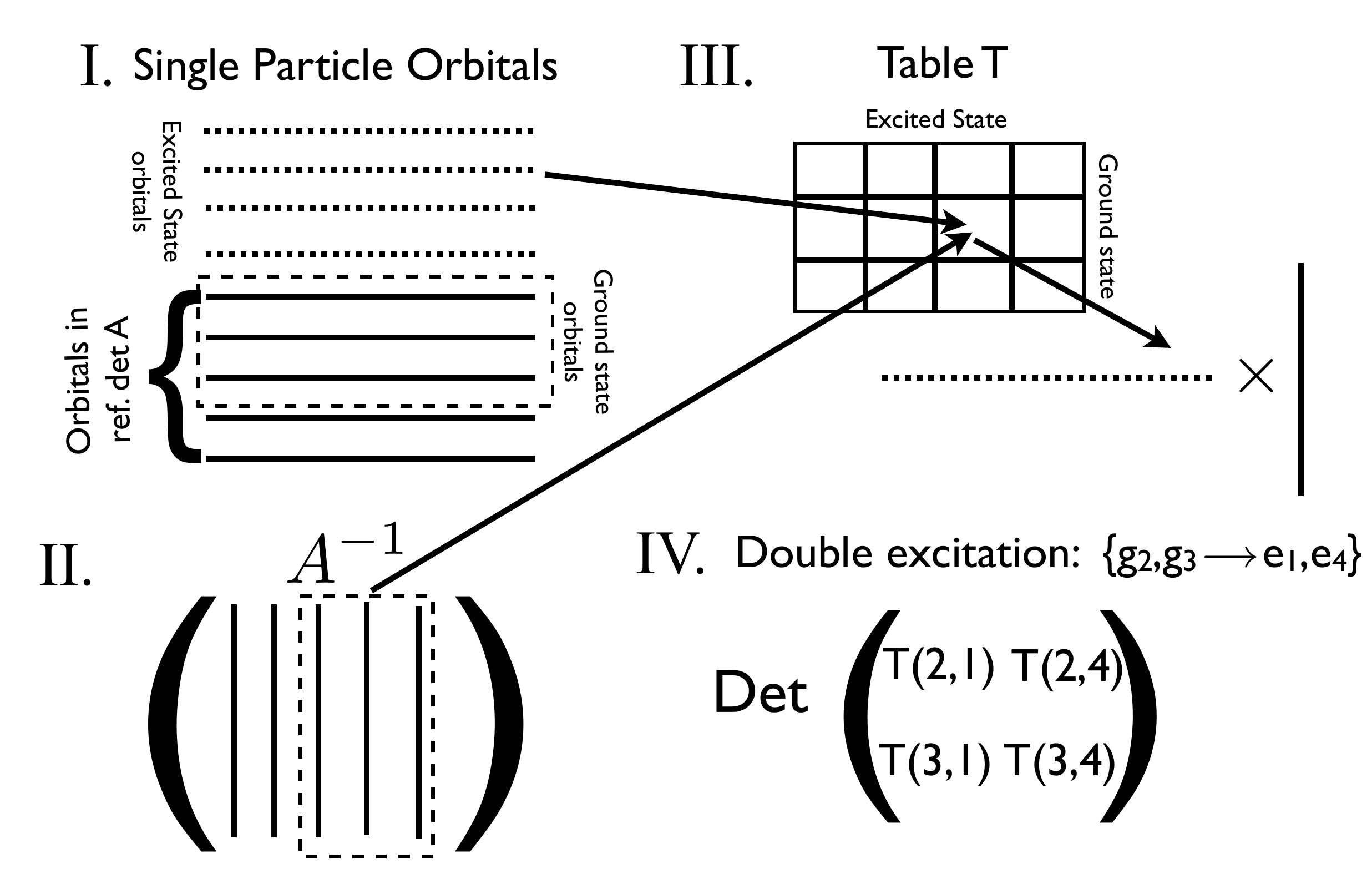}
\label{fig:RatioFigure}
\caption{}

A graphical representation of the key parts of the MDS algorithm.
A table T (step III) is computed from the dot products of orbitals
selected from $A^{-1}$ (step II - stored throughout the simulation)
and the excited state orbitals (step I). Excitations are then read
off (step IV) as small determinants of entries from the table $T.$ 
\end{figure}

In this section we describe a simple, fast procedure for computing
ratios of the multideterminant sum (MDS) evaluated on configurations
$R\equiv\{r_{1}...r_{i}..r_{n}\}$ and $R'\equiv\{r_{1}..r_{i}'...r_{n}\}$
.
The key features of this approach involve only needing to store
a single inverse (that of the reference determinant), as well as being
able to precompute all the necessary information to generate the excitations
at once. Fig.~\ref{fig:RatioFigure} graphically shows the key steps
in this process.

Because a large number of determinant ratios that differ in multiple rows (and 
occassionally columns) need to be computed, 
the basic overall approach will involve using the Sherman Morrison-Woodbury 
formula in evaluating them. We will find that for multiple excitations, though,
the values involved in applying these formulas will be needed over and over again
and so significant savings can be garnered by intelligently caching intermediate results.

For computing these ratios, the following quantities at each Monte Carlo step
(currently located at configuration $R$) must be stored: the inverse
of the reference determinants $M_{\sigma0}^{-1}(R)$ as well as, for
each $k$, the determinant $\det M_{\sigma k}(R)$. At the beginning
of the simulation, the inverse and determinant are calculated from
scratch (although even in the initial step, the determinants can be
calculated faster using some steps of the procedures described below).
Because this only happens once, it does not effect the computational
complexity of our procedure. 

Given that we store (and update) the determinants so that we always
have access to $\det M_{\sigma k}(R)$ and $\det M_{\sigma k}(R')$
it should be clear that to evaluate the determinant ratio \[
\frac{\sum_{k}\alpha_{k}\det M_{\uparrow k}(R')\det M_{\downarrow k}(R')}{\sum_{k}\alpha_{k}\det M_{\uparrow k}(R)\det M_{\downarrow k}(R)}\]
is a straightforward matter of summing and dividing the appropriate
quantities each weighted by $\alpha_{k}$ . In the rest of this section,
then, we focus on the process for updating both $M_{\sigma0}^{-1}$
and (for all $k$) $\det M_{\sigma k}$ as we move from $R\rightarrow R'$ 

In moving from $R\rightarrow R'$ the reference determinants $M_{\sigma0}^{-1}(R')$
needs to be evaluated (recalling we already start with $M_{\sigma0}^{-1}(R)$).
This is exactly the same requirement for the typical Slater-Jastrow
wave-functions and will be accomplished in the canonical manner using
the Sherman-Morrison formula. Because this type of update will be
used throughout this paper in a variety of different contexts, we
will outline the process here for completeness. First note that $ $since
$R$ and $R'$ differ by a single particle, $r_{i}$, this means that
$M_{\sigma0}(R)$ and $M_{\sigma0}(R')$ differ by the change of a
single row. Using the Sherman-Morrison formula one can update the
inverse of a matrix $M_{\sigma0}(R')$ which differs from $M_{\sigma0}(R)$
via a single row by computing

\[
M_{\sigma0}^{-1}(R')=M_{\sigma0}^{-1}(R)-\frac{[M_{\sigma0}^{-1}(R)]ue_{k}^{T}[M_{\sigma0}^{-1}(R)]}{1+e_{k}^{T}M_{\sigma0}^{-1}(R)u}\]
 where $u_{j}=\phi_{j}(r_{i}')-\phi_{j}(r_{i})$ and $e_{k}$ is the
Cartesian basis vector with a 1 at position $k$. In the process of
computing this, it should be noted that the determinant ratio \[
\frac{\det M_{\sigma0}(R')}{\det M_{\sigma0}(R)}=1+e_{k}^{T}M_{\sigma0}^{-1}(R)u\]
 has already been computed from the denominator of our update.  Generically evaluating
this latter quantity involves computing a single dot product and costs $O(n)$ time. Since
$\det M_{\sigma0}(R)$ is already stored, one can compute $\det M_{\sigma0}(R')$
from this ratio. This total operation of updating $M_{\sigma0}^{-1}$
and $\det M_{\sigma0}$ is an $O(n^{2})$ operation.

In addition to updating the reference determinant the new terms $\det M_{\sigma k}(R'),k>0$
need to be evaluated for the new configuration $R'.$ These terms
will be generated by taking the product of $\det M_{\sigma0}(R')$
(computed from the denominator of the inverse updates) and the ratio
$\det M_{\sigma k}(R')/\det M_{\sigma0}(R')$ . Computing these latter
ratios will be the focus of the rest of this section. 

It should be noted that a naive approach to computing each ratio $\det M_{\sigma k}(R')/\det M_{\sigma k}(R')$
involves storing $M_{\sigma k}^{-1}(R)$ for each $k$ and updating
it as we have done for the reference determinant. If we let $n_{e}$
be the total number of determinants this gives us $O(n_{e}n^{2})$
time per step making the use of a MDS cost
$n_{e}$ times a single determinant. This is a significant cost and
consequently instead of computing the ratios $\det M_{\sigma k}(R')/\det M_{\sigma k}(R)$
in this manner, we will do something more sophisticated. 

Recall that in computing the series of ratios $\det M_{\sigma k}(R')/\det M_{\sigma0}(R')$,
we start with the inverse matrix $M_{\sigma0}^{-1}(R')$ where $M_{\sigma k}$
and $M_{\sigma0}$ differ by $s$ columns (and hence is an excitation
of $s$ particle-hole pairs). The general approach for computing these
quantities will be as follows: we fill a table of elements and then
compute the desired ratios by evaluating determinants of small $(s\times s)$
matrices whose elements are taken from this table. Each element $(i,j)$
in the table will be generated via the dot product of the single particle
orbitals $\phi_{i}(R')$ with a column of the matrix inverse $[M_{\sigma0}^{-1}(R')]_{j}$.

In more detail, let $\{g_{1}...g_{k}\}$ be the set of all orbitals
in the reference matrix $M_{0}(R')$ that must be replaced by orbitals
$\{e_{1}...e_{m}\}$ to produce all the excitation matrices $M_{1}(R')...M_{ne}(R')$
in the MDS. Additionally let $\{g_{1}^{-1}...g_{k}^{-1}\}$ be the
corresponding columns in the inverse matrix $M_{0}^{-1}(R')$. An
example double excitation then might replace $\{g_{1},g_{5}\}\rightarrow\{e_{2},e_{7}\}$
. 

To compute the needed ratios, first, generate a table $T$ of size
$k\times m$ where the $(i,j)$ element of the table is $g_{i}^{-1}\cdot e_{j}.$
 Note that each dot product costs $O(n)$ and one must do this for
$km$ different elements in our table giving us a cost of $O(kmn).$
For each single excitation of the form $e_{i}\rightarrow g_{j}$ read
off the $T(i,j)$ element of the table. Notice that if the MDS spanned
all possible pairs of single particle excitations from $\{g\}\rightarrow\{e\}$
(i.e. there are $n_{s}=km$ single excitations), then every element
computed in the table $T$ is utilized. This is a typical occurence
and means all the computation so far was necessary even if only single
particle excitations are calculated. 

The work required for excitations above the single excitation (double,
triple, etc.) is constant per excitation. For each double excitation
of the form $(e_{i},e_{j})\rightarrow(g_{k},g_{l})$, compute the
determinant of the $2\times2$ matrix \[
\det\left(\begin{array}{cc}
T(i,k) & T(i,l)\\
T(j,k) & T(j,l)\end{array}\right)\]
More generically, to compute an excitation replacing $r$ excitations
with $r$ ground state orbitals, compute the determinant of the $r\times r$
matrix where the rows are labelled by the excitation orbitals, the
columns by the ground state orbitals and the matrix elements selected
from their corresponding orbital pairs in $T$. 

Notice, that computing a $r'$th level excitation now simply involves
reading off $r^{2}$ matrix elements and computing the determinant
of a $r\times r$ matrix ($O(r^{3})$ time). Consquently, this step
takes 

\[
O\left(\sum i^{3}n_{i}\right)\]
time where $n_{i}$ is the number of $i-th$ level excitations. Since
$r$ is almost always bounded by some small value less then 8 (rarely
are octuple excitations considered), this becomes an operation proportional
to the number of excitations $O(n_{e})$ . 

In the situation where we span all possible single excitations, this
gives us a total cost per step (including updating the reference determinant)
of \[
O(n^{2})+O(kmn)+O(n_{e})=O(n^{2})+O(n_{s}n)+O(n_{e})\]
 where $n_{s}$ is the number of single excitations and $n_{e}$ the
total number of determinants. Since the cost per step for evaluating
a Slater-Jastrow wave function scales as $O(n^{2})$ this gives us
an additional cost of $O(n_{s}n)+O(n_{e})$ to use multideterminant
wave functions. Algorithm~\ref{alg:Table1} summarizes the cost of
these different steps.

We can also compute the memory cost for using this approach. We need
to store a $n\times n$ matrix (as in the single determinant case)
to store the inverse $M_{\sigma0}^{-1}$ . We also needs to store the
single particle orbitals evaluated on all the particles. These cost
$n^{2}+(m+n)n$ memory and need to be stored throughout the simulation.
At each step, we also need to produce the table $T$ of size $km=n_{s}$
and store the values of the determinant ratios ($\sim n_{e}$). This
gives a complete cost in memory of $(n^{2}+(m+n)n+n_{s}+n_{e})$.
We note that the primary cost in terms of memory is dominated by storing
the single particle orbitals. As is often the case, there is some
tradeoff between memory and computational complexity and the total
memory could be decreased at the cost of some recomputation. 

\begin{algorithm}
\caption{Fast MultiDeterminant when moving from $R\rightarrow R'$}
\label{alg:Table1}
\begin{enumerate}
\item Update the reference determinants inverse $M_{\sigma0}^{-1}(R')$
using the Sherman-Morisson formula. Cost: $O(n^{2})$ 
\item Update the single particle orbitals of the excited states $\{e_{1}...e_{m}\}$.
Cost: $O(mn)$ 
\item For all ground state orbitals $\{g_{1},g_{2}...g_{k}\}$ whose corresponding
row in $M_{\sigma0}^{-1}(R')$ is $\{g_{1}^{-1},g_{2}^{-1}...g_{k}^{-1}\}$
and all excited states $\{e_{1}...e_{m}\}$ compute a $k\times m$
size table $T$ whose $(i,j)$ element is $g_{i}^{-1}\cdot e_{k}$
. This can be done by $km$ dot products or a single matrix multiplication.
Cost: $O(kmn)=O(n_{s}n)$ 
\item To evaluate the ratio of $\Psi(\{g\}\rightarrow\{e\})/\Psi(\{g\})$
, compute the determinant of a matrix whose elements are selected
from $T$ corresponding to the rows and columns that make up all pairs
in $\{g\}$ and \{e\} 
Cost: $O(n_e)$
\end{enumerate}

\end{algorithm}

\section{Gradients and Laplacians}

In FNDMC, in addition to computing ratios, the terms \[
\frac{\nabla_{i}\Psi}{\Psi}\]
and 

\[
\frac{\nabla_{i}^{2}\Psi}{\Psi}\]
 also need to be evaluated. The gradients must be evaluated at each
step while the laplacian is required for evaluating the kinetic energy
and can be evaluated at each step or after one sweep (i.e. a step
of each particle). In the following section we will write $\nabla_{i}^{\alpha}$
to notate one of $(\nabla_{i}^{x},\nabla_{i}^{y},\nabla_{i}^{z},\nabla^{2})$
and will use the term gradients as shorthand for gradients and laplacian.
We assume throughout the rest of this section that we are only working
with one specific particle $i$ . 

As calculating the Jastrow's can be done in the standard way, we again
focus on the gradients of the multi-determinant term. Such terms look
like \begin{equation}
\frac{\nabla_{i}^{\alpha}\sum_{k}\alpha_{k}\det M_{\uparrow k} \det M_{\downarrow k}}{\sum_{k}\alpha_{k}\det M_{\uparrow k} \det M_{\downarrow k}}\label{eq:gradMDS}\end{equation}

The denominator of eqn. \ref{eq:gradMDS} shows up in evaluating the
ratios, and, as such, the machinery necessary to compute this has
already been developed. To compute the numerator, we evaluate the
quantities $\nabla_{i}^{\alpha}\det M_{\sigma k}$ (from which the
numerator is then easily calculable) by evaluating \begin{equation}
\frac{\nabla_{i}^{\alpha}\det M_{\sigma k}}{\det M_{\sigma0}}\label{eq:gradDet0}\end{equation}

\begin{figure}
\includegraphics[scale=0.5]{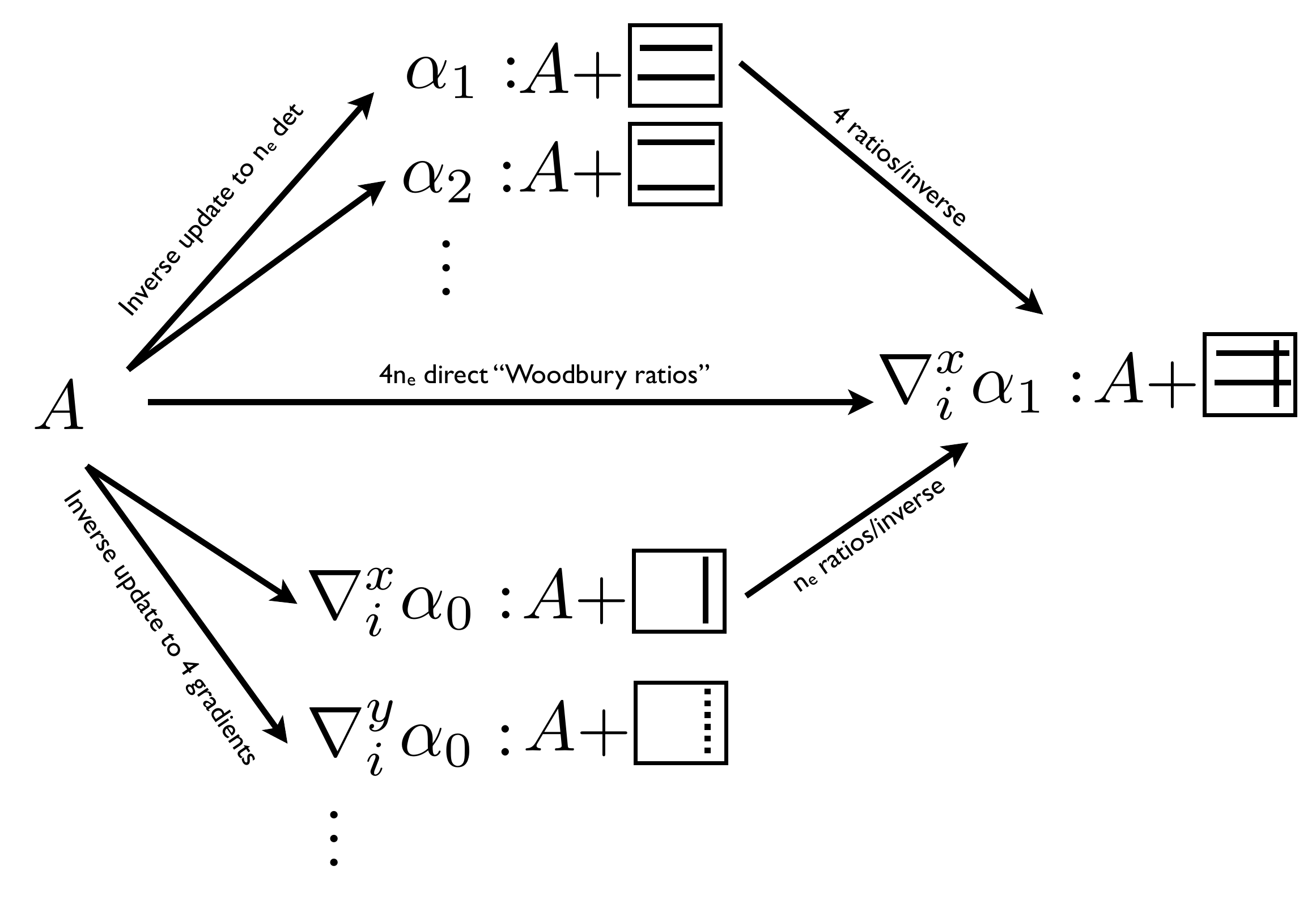}
\label{fig:Deriv}
\caption{}

A graphical representation of the different approaches to computing
the derivative of the matrix that differs by multiple rows and 1 column.
The most efficient path (in most regimes) is to follow the lower one.
\end{figure}

For a given particle $i$ and each of the four $\nabla_{i}^{\alpha}$
terms, eqn. \ref{eq:gradDet0} can be written as the determinant ratio
between two matrices: $M_{\sigma 0}$ and a matrix generated by replacing
$s$-columns (for a $s$ particle excitation) and 1 row (corresponding
to particle $i$) from $M_{\sigma k}$. Because the two matrices differ
in \emph{both} rows and columns, the typical approaches for computing
ratios can not be straightforwardly applied. Instead there are three
possible approaches for dealing with this. Fig.~\ref{fig:Deriv} outlines
these three options.

\subsection{Updating the gradients first}

Recall that we start our evaluation with the reference matrix inverse
$M_{\sigma0}^{-1}$ and need to compute the ratio of two determinants
that differ in both rows and columns. We will compute the ratio of
these two determinants in two steps by working with an intermediate
matrix for each of the four $\alpha$. This intermediate matrix which
we notate as $M_{\sigma0}^{\alpha}$ corresponds to replacing the
elements of the $i'$th row of $M_{\sigma0}$ whose $j'$th element
currently is $\phi_{j}(r_{i}$) with the values (for their $j$'th
element) corresponding to $\nabla_{i}^{\alpha}\phi_{j}(r_{i})$ .
It should be noted that \[
\frac{\det M_{\sigma0}^{\alpha}}{\det M_{\sigma0}}=\frac{\nabla_{i}^{\alpha}\det M_{\sigma0}}{\det M_{\sigma0}}\]
 For each of these 4 matrices, we explicitly compute its inverse
$[M_{\sigma0}^{\alpha}]^{-1}$ using the Sherman Morrison formula.
This costs $O(n^{2})$ work. Recall, as part of this computation,
this also gives us the ratio $\det M_{\sigma0}^{\alpha}/\det M_{\sigma0}$
. Notice that, computing $\det M_{\sigma k}^{\alpha}/\det M_{\sigma0}^{\alpha}$
for a given $k$ would then be sufficient to compute eqn. \ref{eq:gradDet0}
for that $k$. These two matrices, though, differ from each other
by the same series of particle-hole excitations that we worked with
in the ratios. In order to compute $\det M_{\sigma k}^{\alpha}/\det M_{\sigma0}^{\alpha}$
for all $k$ we will do exactly the same thing as we did for the determinant
ratios! The only difference in this case is when building the table
$T$, each element of the sets $\{e\}$ have their $i'th$ value replaced
by the gradient of that value and the elements in the set $\{g\}^{-1}$
now differ because the reference inverse is different. It is clear
then that the cost of this operation is four times the cost of evaluating
a determinant ratio. Therefore, computing all the gradients and laplacians
for a single particle costs $4(n^{2}+nn_{s}+n_{e})$ . This means
that computing the gradients and laplacians twice per step (the amount
required for FNDMC) is only a constant factor more costly then computing
the determinant ratios themselves. Although this is worse then the
single determinant case (where the ratios can all be computed without
updating any inverses) the total computational complexity is still
well controlled. 

\subsection{Updating each determinant first}

As an alternative approach, instead of producing an updated inverse
for each of the four $\nabla_{i}^{\alpha}$ we instead could have
produced an updated inverse for each multi-determinant matrix and
then do row-ratios with respect to the four $\nabla_{i}^{\alpha}$
we need. For the computation of the gradient of a single particle
this is always an inferior approach. If the gradient is being computed
for all particles simultaneously, then in this alternative approach,
the updated inverse needs only be computed once and so its cost can
be amortized over all particles. (No such savings for doing all particles
simultaneously can be garnered in the table method). Working out the
cost for computing all the terms (i.e. one sweep) gives a cost of $O(n^{2}n_{e}).$
This is computationally superior in the situation where $n^2n_{e}<n^{3}$.
We expect such situations to be rare but in these cases and when all
particles need to be calculated simultaneously, this latter approach
scales asymptotically better.

\subsection{Direct Application of the Generalized Matrix Determinant Lemma}

The generalized matrix determinant lemma states \[
\det(A+UV^{T})/\det A=\det(I+V^{T}A^{-1}U)\]
There exists choices of $U$ and $V$ (taking matrices of size $(s+1)\times n$)
that allow $A+UV^{T}$ and $A$ to differ by $s$ columns and $1$
row.  Then, for each $i$ and each $\alpha,$ one could directly apply
this lemma. An upper bound on this cost per $(i,\alpha)$ is $(4n^{2}+16n)$
(the cost can be brought down somewhat because some of the matrix
vector multiplications involved are against cartesian basis vectors
and hence can be computed quickly). Performing this computation for
the gradients and laplacians of all the excitations will then take
$16n^{2}n_{e}$. This is prohibitively expensive and not the correct
approach when computing the gradients or laplacians for all the particles.
Nonetheless, it might be a useful method to have in the rare case
where a single term needs to be computed. Additionally, this has the
added benefit of not having numerical instability problems that can
potentially crop up in other approaches.

\section{Additional Considerations}

In this section, we present a couple of additional considerations
that should be considered when implementing the Table method. Specifically
these involve efficiency considerations, numerical instabilities,
and parallelization.

To begin with, the dominating effort in computing determinant ratios
involves the computation of dot products or rows/columns of matrices.
The performance of the entire algorithm will depend heavily on optimizing
the application of these dot products. Specific care should be taken
to ensure that the dot products are taken so the fastest index is
being looped over. Additionally, in a number of cases, these
dot products can be chunked together into a single large matrix multiplication
step (as opposed to a series of dot products). For example, in computing
the elements of the table, instead of computing all pairs of dot products
between the sets of $\{g\}^{-1}$ and $\{e\}$ one can instead compute
all these dot products simultaneously as a matrix multiplication between
two matrices whose rows/columns are respectively the vectors of $\{g\}^{-1}$and
$\{e\}$. Not only will this result in significantly better cache
performance, and an ability to take advantage of highly-tuned matrix
multiplication routines, but formally \footnote{Matrix multiplication is known to
scale theoretically better then $O(n^3)$ (i.e. at least $O(n^{2.376})$) but these asymptotically
faster algorithms are rarely useful in practice.}  can also be asymptotically
faster. 

Concerning numerical instabilities, there are two places in this algorithm
where such instabilities can crop up. (Although these are 
both theoretically possible, we never find them to be a problem in practice
and so have not yet been forced to implement the suggested methods to avoid them).
To begin with, the inverse of
the reference determinant $M_{0}$ could become (close to) singular.
When evaluating wave functions of the Slater-Jastrow type this is
never a problem because if $M_{0}$ becomes (close to) singular, the
probability of moving to that configuration is essentially 0 and so
the move is rejected. This can not be ensured in the case of multiSlater-Jastrow.
It could be the value of the reference matrix goes to 0 but since
the other determinants have a non-zero value, the move itself would
be accepted. This problem can be dealt with by temporarily switching to
another of the determinants in our expansion as the reference determinant.
This increases the cost slightly as the the size of all the excitations
may potentially increase by the number of particle-hole excitations
in the new reference determinant, but this new reference determinant
need only be used for small portions of the simulation. 

The second area where numerical instabilities might appear is in the
calculation of the gradients. Again it is possible in the inverse
update step, that the matrices end up being nearly singular. Of course,
again one could potentially find an appropriate new reference determinant.
Because an update step is not strictly required in this case, though,
another option exists. Instead of updating an inverse at all, for
each multi-determinant, we can evaluate the ratio with respect to both
rows and columns simultaneously using the generalized matrix-determinant
lemma (as described in section IV-C). $ $ This can be done in time
$O(n^{2})+O(nn_{e})$ (to achieve even this scaling, one must be careful
not to perform identical matrix multiplications twice). We note that
this is potentially significantly more expensive then the table method
($nn_{e}$ instead of $nn_{s}$), but needs only to be done upon seeing
numerical instabilities.

Finally, it should be pointed out that the algorithms described in
this paper can be easily parallelized especially over many cores
on a single machine. The table of elements can be easily broken up
into chunks each of which can be computed by separate cores. Moreover,
once this table has been produced, reading and computing the tiny
determinants can be done in parallel. Consequently, using this approach
we believe that the evaluation of these multideterminant terms can
scale particularly well.

\section{Water Molecule~}

In this section, we benchmark our approach on the all electron water molecule at the
equilibrium geometry r$_{OH}$ = 0.9572 \AA, $\theta_{HOH}$ = 104.52$^{\circ}$. 
There are accurate results for this molecule and it has received considerable attention 
from the QMC community in the past \citep{Sorella04,Gurtubay07,Needs06,Luchow00}.
Since we do not employ orbital optimization \citep{Umrigar08} in our QMC calculations,  
we rely on high level quantum chemistry methods to produce a good set of molecular 
orbitals for our multideterminant expansion. All quantum chemistry calculations were
performed with the GAMESS code \citep{Gamess}, using the Roos augmented triple zeta 
ANO Gaussian basis set \citep{Widmark1990,EMSL,Feller96}. To obtain the molecular orbitals, we start
with a Complete Active Space Self-Consistent Field (CASSCF) calculation including 
all 10 electrons in 8 orbitals. The resulting orbitals are used in a Multi Configuration Self-Consistent Field (MCSCF) calculation
including all single and double excitations from the CAS involving the next 44 orbitals; this
is known as second-order CI (SOCI) in the chemistry community. We used the natural orbitals
of the converged MCSCF calculation and choose the configurations in our expansion by applying a 
cutoff to the SOCI wave function. 
We found this approach to balance the complexity of the quantum
chemistry calculation with the quality of the resulting set of orbitals and configurations. 
Instead of
using determinants directly in our calculations, we use configuration state functions (CSF) which
are spin and space adapted linear combination of Slater determinants. This eliminates redundant
variational parameters from the wave function and facilitates the optimization process. To this 
multideterminant expansion we add a Jastrow factor that contains 1,2, and 3 body terms and optimize
all the variational parameters simultaneously (including the nonlinear Jastrow coefficients and 
the linear CSF coefficients) by energy minimization
based  on a variant of the linear method of C. Umrigar, et. al. \citep{Umrigar08}. 

Figure 3 shows the speedup (ratio of the time required between the base and Table methods) obtained with the Table method in the evaluation of the ratios and gradients of the list of determinants involved in the construction of the wavefunction during a VMC move. The blue line does not include the time spent performing the inverse updates in the canonical algorithm (nor the inverse update of the reference determinant in the Table algorithm). For the comparisons including inverse updates two time steps are shown, giving approximately a 50$\%$ and 100$\%$ acceptance rate in the VMC calculation. These cases represent the two typical limits encounter in QMC. Although the exact speedup depends on the details of the particular problem, with a few thousand determinants we get speedups on the order of 25. For fixed $n$ (and fixed or small $n_s$), both the 
canonical and Table algorithms scale as $O(n_e)$ and therefore for large $n_e$ we expect speedup to be a constant
factor as we see in Figure 2. Because the prefactor before this constant goes as $n^2$ for the canonical algorithm 
and constant for the Table method, we expect the speedup to grow for larger systems.  Oscillations seen before 
reaching this assymptotic limit, is caused by implementation details, e.g. cache usage, locality of 
references in memory, etc. These timings are not meant to be a reflection of the operation counts involved in the algorithm, but merely an example of the increased efficiency of our implementation in QMCPACK.     
\begin{figure}[h]
\includegraphics{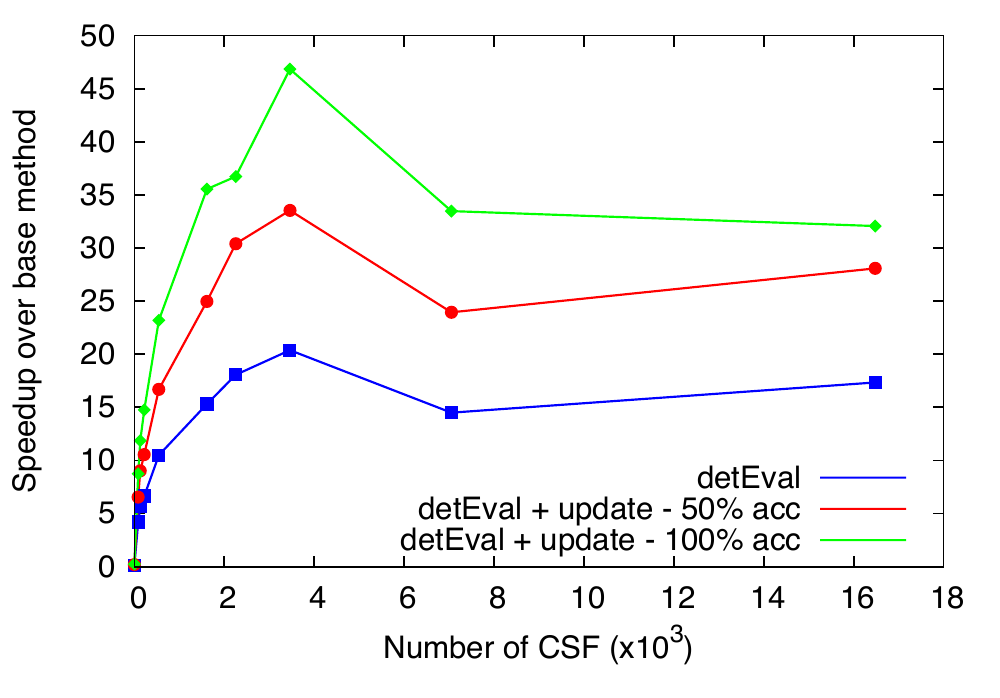}
\caption{ Speedup obtained with the Table method in the evaluation of the ratios and gradients of the list of determinants involved in the construction of the wavefunction during a VMC move. The blue line only includes the calculation of the ratios and gradients, while the other lines also include the inverse update. The red and green lines use time steps that produce acceptance rates of 50$\%$ and 100$\%$ respectively.}
\label{Fig:speedup}
\end{figure}


Figure 4 shows our VMC and DMC total energies of the water molecule as a function of the sum of the squares of the initial CSF coefficients (those produced by the MCSCF calculation). The latter is related to the fraction of important configurations (according to the MCSCF calculation) included in the expansion, as we approach unity we include all the determinants in the SOCI wavefunction. Our results not only show the ability of this wavefunction to recover large amounts of correlation, but also the systematic improvement of the energy as more configurations are included. Notice that a stable and reliable optimization method is needed in order to recover the maximum amount of correlation, the details of our optimization algorithm will be discussed in a future publication \cite{Mcminis11}. Table 1 shows a comparison of our calculations with a selection of previous results. Our best wavefunction (including 3461 CSF) recovers 99.7$\%$ of the correlation energy, with an error of 1.2 mHa.  It should be noted that the energy computed in DMC with the determinants selected from the SOCI wavefunction is significantly better then the energy that comes out of the SOCI calculation directly, $E_{SOCI}$ = -76.3619395110.

\begin{figure}[htdp]
\includegraphics{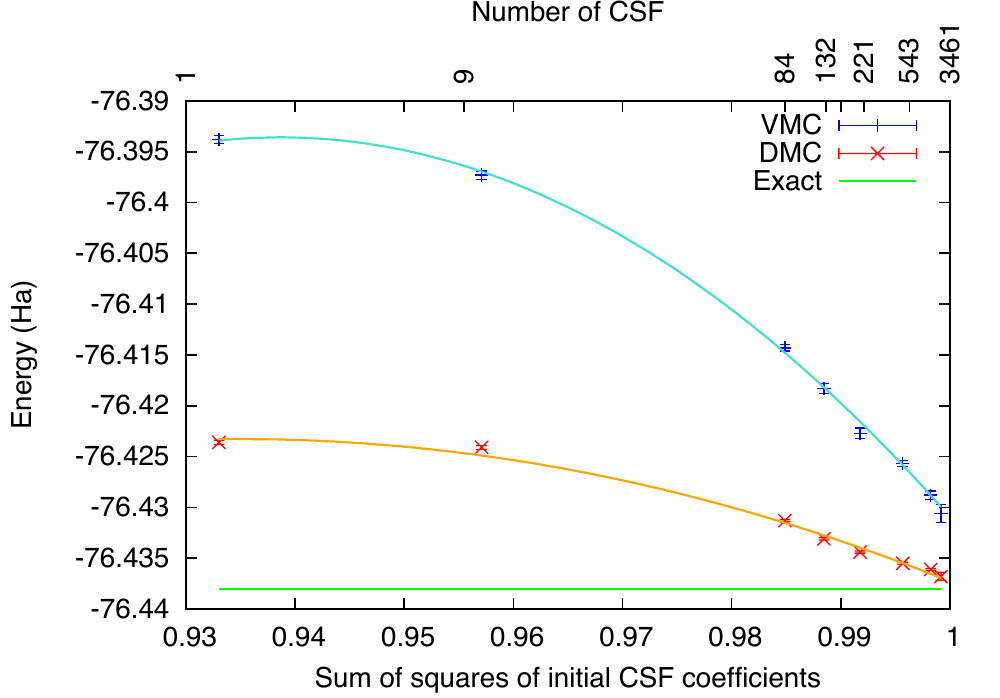}
\caption{Total energy of the water molecule as a function of the sum of squares of the initial CSF coefficients.} 
\label{Fig:H2OEnergy}
\end{figure}

\begin{table}[htdp]
\begin{center}
\begin{tabular}{c c c c}
\hline
Method    &    Total Energy (Ha)  &  Ref.   \\
\hline
 VMC-SingleDet   & -76.3938(4) & This work. \\
 VMC-MSDJ   &  -76.4306(9) &  This work. \\ 
 VMC-SJB   &  -76.4034(2)  &  \cite{Gurtubay07}     \\
 \hline
 DMC-SingleDet   &  -76.4236(2) & This work. \\
 DMC-MSDJ   & -76.4368(4) & This work. \\
 DMC-B3LYP  & -76.4230(1)  &  \cite{Gurtubay07} \\
 DMC-SJB   &  -76.42830(5)  &   \cite{Gurtubay07} \\
 DMC-AGP   &  -76.4175(4)  &  \cite{Sorella04} \\
 DMC-PNO-CI   &   -76.429(1)  &  \cite{Luchow00}  \\
 CCSD(T)-R12  &  -76.4373   &  \cite{Muller97} \\
 CEEIS    &    -76.4390(4)   &  \cite{Bytautas06} \\
 Exact    &    -76.438  &  \cite{Feller86} \\
 \hline
\end{tabular}
\end{center}
\label{Table:H2O_energies}
\caption{Comparison of results on the water molecule. SingleDet refers to our calculations employing only a single determinant and MSDJ refers to our multiSlater-Jastrow wavefunction.  }
\end{table}%

\section{Conclusions}

Although multideterminant expansions were introduced in QMC more 
than a decade ago, their use in molecular problems has been uncommon
and the number of configurations typically included in calculations is 
rather small.  In this work we have described an algorithm that allows for 
their efficient evaluation. Specifically
we focus on the steps of computing wave function ratios and evaluating
gradients and laplacians of these wave functions. Beyond the typical
inverse update of a reference matrix that is required for just the
Slater-Jastrow ansatz, the algorithm for evaluating ratios involves
two key steps: computing a table (whose elements are often all computable
by a single matrix multiplication) and then reading off ratios from
elements (or determinants of a few elements) of this table. Gradients
and laplacians can be computed in the same way as ratios where the
reference matrix being used is the matrix where a row is replaced
by its gradient (respectively laplacian). This procedure allows the
use of thousands of determinants for a cost that is only a few times
the cost of a single determinant. We test this empirically on a water
dimer getting good energies at a reasonable computational cost. 
This algorithm will open up the possibility of retrieving a
significant percent of the correlation energy in a variety of strongly
correlated systems. 

%
%
%

\begin{acknowledgments}
We would like to thank Ken Esler for useful conversations and 
Cyrus Umrigar for carefully reading and commenting on the manuscript.
The work at Rice University was supported by the Department of Energy
(DE-FG02-04ER15523) and the Welch Foundation (C-0036).
This work was performed in part under the auspices of the US DOE by LLNL
under Contract DE-AC52-07NA27344. 
The work at UI was supported by the National Science Foundation under No.
0904572 and EFRC - Center for Defect Physics sponsored by the US DOE, Office of Basic Energy Sciences.

\end{acknowledgments}

\bibliographystyle{apsrev}
\bibliography{MultiDets6}

\end{document}